\documentclass[12pt, a4paper]{article}  
\usepackage{lscape}
\usepackage{typearea}
\typearea{12}
\input{epsf.tex}
\begin{document}
\setlength{\parskip}{2ex}
\def\bea{\begin{eqnarray}}
\def\eea{\end{eqnarray}}
\def\6{\partial} \def\a{\alpha} \def\b{\beta}
\def\c{\gamma} \def\d{\delta} \def\ve{\varepsilon} \def\e{\epsilon}
\def\z{\zeta} \def\h{\eta} \def\th{\theta}
\def\vta{\vartheta} \def\k{\kappa} \def\l{\lambda}
\def\m{\mu} \def\n{\nu} \def\x{\xi} \def\p{\pi}
\def\r{\rho} \def\s{\sigma} \def\t{\tau}
\def\Ph{\phi} \def\ph{\varphi} \def\ps{\psi}
\def\o{\omega} \def\G{\Gamma} \def\D{\Delta}
\def\Th{\Theta} 
\def\Lam{\Lambda} 
\def\S{\Sigma}
\def\PH{\Phi} \def\Ps{\Psi} \def\O{\Omega}
\def\sm{\small} \def\la{\large} \def\La{\Large}
\def\LA{\LARGE} \def\hu{\huge} \def\Hu{\Huge}
\def\ti{\tilde} \def\wti{\widetilde}
\def\non{\nonumber\\}
\def\ll{\Longleftarrow}
\def\lr{\Longrightarrow}
\def\semidirect{\;{\rlap{$\subset$}\times}\;}
\def\stareq{\ {\buildrel{*}\over =}\ }
\def\xt{{\tilde x}}
\def\quer{\!\!\!\!\!\!\!\nearrow}  
\def\Reel{R}
\def\blacksquare{ \vbox{ {\hrule height 1pt width 2pt depth 1pt } } }
\def\inner{\rfloor }
\def\eqa{\eqno(\z )}\def\vta{\vartheta}
\def\starequal{\buildrel \ast\over =}
\def \bbbone{1}
\def\aPsi{\overline\Psi}
\def\hodge {{}^\star\!}

\newcount\secno
\secno=0
\newcount\susecno
\newcount\fmno\def\z{\global\advance\fmno by 1 \the\secno.
                       \the\susecno.\the\fmno}
\def\section#1{\global\advance\secno by 1
                \susecno=0 \fmno=0
                \centerline{\bf \the\secno. #1}\par}
\def\subsection#1{\medbreak\global\advance\susecno by 1
                  \fmno=0
       \noindent{\the\secno.\the\susecno. {\it #1}}\noindent}


\newcount\refno
\refno=1
\def\y{\the\refno}
\def\myfoot#1{\footnote{$^{(\y)}$}{#1}
                 \advance\refno by 1}


\def\newref{\vskip 0.3pc 
            \hangindent=2pc
            \hangafter=1
            \noindent}

\def\neq{\hbox{$\,$=\kern-6.5pt /$\,$}}


\def\asteq{\buildrel \ast \over =}


\font\fbg=cmmib10\def\clom{\hbox{\fbg\char33}}

\def\semidirect{\;{\rlap{$\supset$}\times}\;}


\newcount\secno
\secno=0
\newcount\fmno\def\z{\global\advance\fmno by 1 \the\secno.
                       \the\fmno}
\def\sectio#1{\medbreak\global\advance\secno by 1
                  \fmno=0
       \noindent{\the\secno. {\it #1}}\noindent}

\null
\bigskip\bigskip\bigskip\centerline{\bf Stress and Hyperstress as Fundamental 
Concepts}
\centerline{{\bf in Continuum Mechanics and in Relativistic Field 
Theory}\footnote{Modified version of an invited lecture given in honor of A.\
Signorini, cf.{\it Advances in Modern Continuum Dynamics},
International Conference in Memory of Antonio Signorini, Isola d'Elba,
June 1991. G. Ferrarese, ed. (Pitagora Editrice, Bologna 1993)
pp.1--32.}}

\bigskip 
\centerline{by}\medskip
\centerline {Frank Gronwald and Friedrich W. Hehl}
\centerline{Institute for Theoretical Physics, University of Cologne}
\centerline{D--50923 K\"oln, Germany}\bigskip\bigskip\bigskip
\centerline{\bf Abstract}
\medskip
The notions of stress and hyperstress are anchored in  
3-dimensional continuum mechanics. Within the framework of the 
4-dimensional spacetime continuum, stress and hyperstress translate  
into the energy-momentum and the hypermomentum current, respectively. 
These currents describe the inertial properties of classical matter 
fields in relativistic field theory. The hypermomentum current 
can be split into spin, dilation, and shear current. We discuss 
the conservation laws of momentum and hypermomentum and point out
under which conditions the momentum current becomes symmetric.

\bigskip\bigskip
\centerline{\bf Contents}
\begin{description}
\item{1.} Introduction and Summary
\item{2.} Force and Hyperforce 
\item{3.} Stress and Hyperstress as $(n-1)$--Forms
\item{4.} Hypermomentum Current 
\item{5.} Dirac Field and its Momentum and Spin Currents 
\item{6.} Quadrupole Excitations, Signorini's Mean Stresses
\item{7.} Conservation Laws for Momentum and Hypermomentum
\item{} Acknowledgments
\item{} Appendix
\item{} References
\end{description}
\vfill\eject

\bigskip

\sectio{\bf Introduction and Summary} 

\medskip
Historically, continuum mechanics had a far-reaching influence on the 
development of relativistic field theory. The {\it Maxwell stresses} 
${\buildrel{\rm Max}\over\sigma}{}^{ab}$ of 
electrodynamics, for example, were conceived by Maxwell as ``being the 
same in kind with those familiar to engineers'' ($a,b=1,2,3$). 
Accordingly, the stress 
concept of Euler and Cauchy, which originally arose in the context of 
bending of beams, of pressure in fluids, and the like, found its way into 
electrodynamics. In this framework, it describes the mechanical state of 
the vacuum (`aether'). Later, the Maxwell stress was interpreted 
by Lorentz as {\it momentum flux density} of the electromagnetic
field and experimentally verified, as `light pressure', 
by Lebedew and Gerlach. 

The next step was taken by Minkowski: He introduced
the 4-dimensional energy-momentum tensor 
${\buildrel{\rm Max}\over\sigma}{}^{ij}$ 
of the Maxwell field, which 
unites into a single quantity the energy density, the energy 
flux density, the momentum density, and the Maxwell 
stress of the electromagnetic field. This 
symmetric 2nd rank tensor ${\buildrel{\rm Max}\over\sigma}{}^{ij}$, 
with coordinate indices $i,j=0,1,2,3$, 
is of fundamental importance, since it acts as one of the source terms 
on the right hand side of Einstein's field equation of gravity. 
Appropriately interpreted, ${\buildrel{\rm Max}\over\sigma}{}^{ij}$ 
represents the (4-)momentum current of the Maxwell field and, 
integrated over a 3-dimensional (spacelike) hypersurface, yields the 
(4-)momentum contained therein. Therefore 
${\buildrel{\rm Max}\over\sigma}{}^{ij}$ should be understood as 
a covector-valued 3-form ($\wedge =$ exterior product)
$${\buildrel{\rm Max}\over\sigma}{}_\alpha=
{1\over 3!}\,{\buildrel{\rm Max}\over\sigma}{}_{ijk\,\alpha}\,
dx^i\wedge dx^j\wedge dx^k\,,\eqa$$  
the covector index $\a$ indicating the momentum direction and the 
form indices $i,j,k$ characterizing the hypersurface under 
consideration. 
Componentwise, we will later find the relation between the 2nd rank 
tensor and the corresponding covector-valued 3-form according to
$${\buildrel{\rm Max}\over\sigma}{}^i{}_\a={1\over 3!}\,\eta^{ijkl}\,
{\buildrel{\rm Max}\over\sigma}{}_{jkl\,\alpha}\,,\eqa$$ 
with $\eta^{ijkl}$ as the totally antisymmetric unit tensor. 

Thus, from the 3-dimensional `engineering stress' \`a la Maxwell 
${\buildrel{\rm Max}\over\sigma}{}^{ab}$ we were
led, in (1.1), to the 4-dimensional momentum current 
${\buildrel{\rm Max}\over\sigma}{}_\alpha$, a kind of a 4-dimensional 
stress, which is able to deform the Minkowski spacetime, via gravity, 
to the Riemannian one of Einstein's theory.

On the other hand, the notion of `engineering' stress arises if one 
smears a force $f_a$ over a 2-dimensional area element. Hence stress carries 
the dimension {\it force/length}$^2$. Whereas in statics and in classical  
point particle mechanics {\it force} represents the fundamental 
dynamical quantity, in continuum mechanics distributed force, that is 
{\it stress}, plays the role of the primary dynamical agent. 
Accordingly, the notions `force' -- `stress' -- `momentum current' 
extend from ordinary statics to general relativity, and we will 
collect the corresponding quantities and formulas governing them 
in a concise way. 

\begin{figure}[htb]
  \epsfbox[-10 0 500 270]{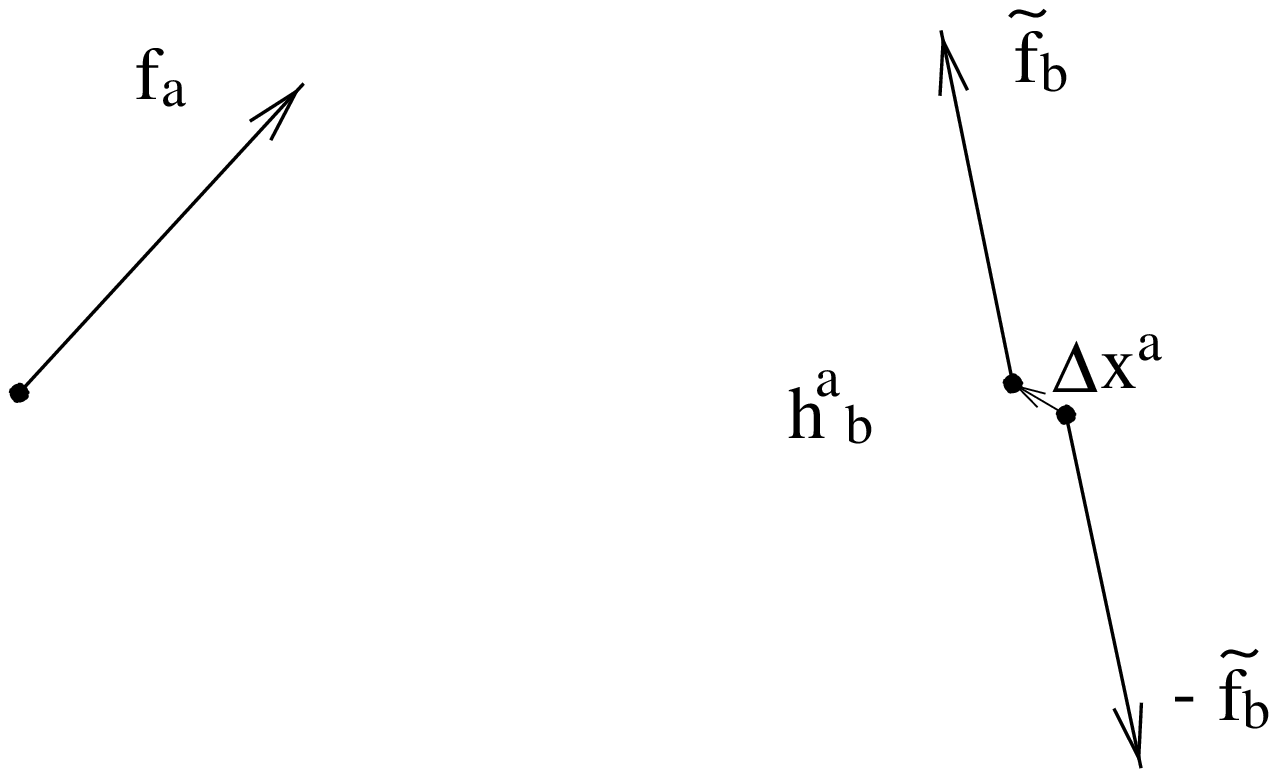} 
\label{forcedef}
\end{figure}

\noindent{\bf Fig.1.} Force: Arrow symbolizing the force 1-form 
$f=f_a\,dx^a$.

\noindent{\bf Fig.2.} Hyperforce: Two opposite arrows displaced 
with respect to each 
other and symbolizing a hyper\-force 1-form $h^a=h^a{}_b\,dx^b$ with 
$h^a{}_b=\lim\Delta x^a\tilde f_b$. Only after a suitable limiting 
transition with $\Delta x^a\rightarrow 0$ and 
$\tilde f_b\rightarrow\infty$, the double force becomes the 
hyperforce $h^a$.

In ordinary 3-dimensional mechanics, besides the force $f_a$, 
see Fig.1, we have the moment $\Delta\vec r\times\vec f$, which is related to 
the rotational motion of the matter configuration considered. 
Appealing to concepts developed in modern continuum mechanics, we 
will generalize the moment right away to the concept of a 
{\it hyperforce}, see Fig.2, the limit of an arbitrary double force
$$h^a{}_b:=\lim\Delta x^a\tilde f_b\,,\eqa$$
the antisymmetric part of which $m^{ab}=\Delta x^{[a}\tilde 
f^{b]}$ is the moment $m_c=\eta_{cab}\,m^{ab}/2$, with 
$\eta_{cab}$ as the totally antisymmetric unit tensor in 3 
dimensions. Its symmetric part $h^{(ab)}$ is new and arises from 
double forces without moment. 

In analogy with the case of a force, we are 
led, starting with statics and ending up with general relativity, 
to the sequence `hyperforce' -- `hyperstress' -- `hypermomentum 
current'. We will postulate in this context that the hyperforce 
represents a new fundamental dynamical quantity which is 
intrinsic and which cannot be resolved in general into a double 
force, that is, the pair `force' $f_a$ and `hyperforce' $h^a{}_b$ 
is considered to lie at the foundations of mechanics. Smearing 
hyperforce $h^a{}_b$ with dimension {\it force}$\times${\it 
length} over a 2-dimensional area element, results in the 
hyperstress (2-form) $\Delta^a{}_b$ with the dimension {\it 
force/length} and, eventually in the 4-dimensional spacetime 
picture, to the hypermomentum current (3-form) $\Delta^\a{}_\b$. 
Its antisymmetric part $\tau^{\a\b}:=\Delta^{[\a\b]}$ is 
well-known as spin current.

In our article we want to support our view that force -- 
hyperforce, stress -- hyperstress, and momentum current -- 
hypermomentum current are fundamental notions which link together 
continuum mechanics and relativistic field theory. In particular, 
the concept of an intrinsic hypermomentum current, which 
takes a central position as intrinsic hyperstress in the 
mechanics of structured media, has been neglected by most 
researchers in relativity theory for no good reason other than 
that Einstein was not aware of it in 1915. We will see that this 
question is closely related to the symmetry of the 
energy-momentum tensor of matter -- another taboo in relativity 
theory.   

In order to fix our position and to give possible challengers the 
opportunity to attack definite statements, we collected the outcome 
of our article in some theses which at the same time summarize our 
results: 

\noindent
{\bf Thesis 1:} Force $f$ (with dimension {\it force}) and hyperforce 
$h^{\alpha}$ (with dimension {\it force}\ $\times${\it length}) 
are basic concepts in the mechanics of structured continua. They 
are scalar- and vector-valued 1-forms, respectively, which occur 
likewise in $n=1,2,3,\,{\rm and}\,\, 4$ dimen\-sions.--

\noindent
{\bf Thesis 2:} Stress $\Sigma_{\alpha}$ (with dimension 
{\it force/length}$^2$) and hyperstress $\Delta^{\alpha}{}_{\beta}$ 
(with dimension {\it force/length}) are basic fieldtheoretic concepts in 
physics. They are covector- and tensor-valued $(n-1)$-forms, 
respectively, which occur likewise in $n=1,2,3,\, {\rm and}\,\, 4$ 
dimensions. In 4 dimensions, 
$\Sigma_{\alpha}$ represents the energy-momentum current and 
$\Delta^{\alpha}{}_{\beta}$ the spin$\oplus$dilation$\oplus$shear 
current of matter.--

With these two theses, the fundamental assumptions are 
laid down. Since we honor Signorini with this Conference, it may be 
appropriate to quote some results which are related to 
one aspect of Signorini's work: 

\noindent
{\bf Thesis 3a:} The theory of Signorini's mean stresses, well developed
for 3 dimensional continuum mechanics, can also be applied to 
the 4-dimensional energy-momentum current.--

\noindent
{\bf Thesis 3b:} In 4 dimensions, suitable components of the time 
derivative of the quadrupole moment of the energy density of 
matter generate, together with the Lo\-rentz generators, 
the Lie-algebra of the 
4-dimensional special linear group $SL(4,R)$.-- 

After this interlude, we come back to the study of stress and hyperstress.
Restricting ourselves to static situations in 1, 2, and 3 dimensions, 
we can write down the equilibrium conditions for stress and hyperstress.
They look the same as the corresponding conservation laws in 4-dimensional 
spacetime. Let $\vta^\a$ be the local 1-form basis, $D$ the exterior 
$GL(4,R)$-covariant derivative, and $g_{\b\c}$ the 
components of the metric, then we have

\noindent
{\bf Thesis 4a:} The balance laws for force and hyperforce in 
n dimensions read:
\bea
   (I) \qquad\qquad\qquad\qquad D\Sigma_{\alpha} &=& \,0\,,\nonumber\\
   (II)\hspace*{0.1cm}\qquad D\Delta^{\alpha}{}_{\beta}+ 
\vartheta^{\alpha}\wedge\Sigma_{\beta}
 &=&\,g_{\beta \gamma}\,\sigma^{\alpha \gamma} \,.\nonumber 
\eea
They are the Noether identities belonging to the group 
$R^n\semidirect GL(n,R)$. On the right hand side of $(II)$, there 
enters a suitably defined symmetric Cauchy stress $\sigma^{\a\c}$. 
The antisymmetric part of $(II)$ represents the conventional balance 
law for moment $D\tau^{\a\b}+\vta^{[\a}\wedge\Sigma^{\b]}=0$. Provided
the intrinsic moment stress $\tau^{\a\b}$ vanishes, the stress becomes
symmetric: $\vta^{[\a}\wedge\Sigma^{\b]}=0$.--

In 3 dimensions, in a dislocated medium, the underlying material 
manifold can be non-Euclidean. Also spacetime, if gravity is allowed for, 
can relax to a non-Minkowskian state ($\rfloor$ = interior product):

\noindent
{\bf Thesis 4b:} Formulated in a non-Euclidean continuum, the 
balance law $(II)$ is left unchanged whereas $(I)$ is modified to:
$$D\Sigma_{\alpha} = (e_{\alpha}\rfloor  
T^{\beta})\wedge\Sigma_{\beta}+ (e_{\alpha}\rfloor 
R_{\beta}{}^{\gamma})\wedge\Delta^{\beta}{}_{\gamma} - {1\over 2}
(e_{\alpha}\rfloor Q_{\beta\gamma})\wedge\sigma^{\beta \gamma}\,.$$  
Here $T^\b:=D\vta^\b$ denotes the torsion of the continuum, $R_\b{}^\c$ its 
curvature, and $Q_{\b\c}:=-Dg_{\b\c}$ its nonmetricity.--

\bigskip\goodbreak
\sectio{\bf Force and Hyperforce}

\medskip
In Newtonian as well as in special-relativistic point particle 
mechanics, the concept of force is fundamental for the formulation 
of the equation of motion. What type of quantity should we assign to 
a force? The notion of force emerged in the context of investigating 
static equilibrium. The typical tool for measuring a force is the 
spring balance (Fig.3). The axis of the spring balance determines a unit 
direction $\hat v:=\vec v/v$. We apply the spring balance in this 
direction and read off a numerical value from its scale. In other 
words, force determined in this way represents a 1-form [6]. 

This result is in accord with the special case of a conservative force.
Then the force 1-form is exact, $f=-d\phi\,$, with 
$\phi$ as the potential 0-form. Moreover, in Lagrangian mechanics, the 
force $f_a:=\partial L/\partial x^a$  is also a 1-form by definition. 
Consequently, the 1-form character of force is well-established. 
Nevertheless, in our figures, for illustrative purposes, we drew the 
force as an arrow, which seems more intuitive than using two parallel 
planes with a specified direction, see Burke [1]. 

\pagebreak

\vspace*{-2cm}
\begin{figure}[htb]
  \epsfbox[-10 0 500 270]{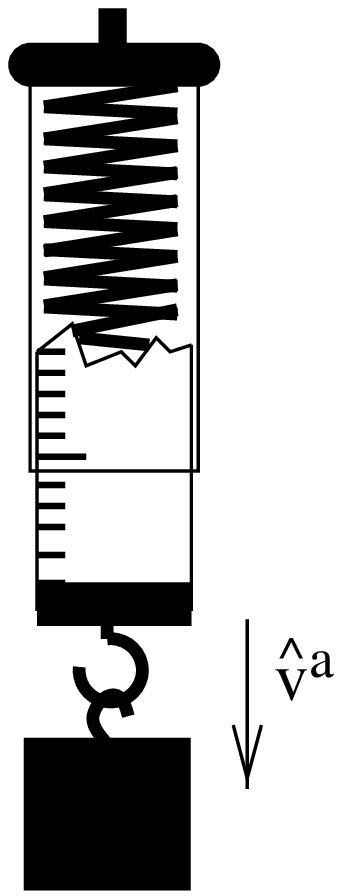} 
\label{force}
\end{figure}
\noindent{\bf Fig.3.} Spring balance for measuring force: 
The axis of the spring balance determines a direction (vector), whereas 
the reading yields a scalar value. Consequently force $f$ is a 1-form. 
If a vector $\hat v$ is applied to $f$, it yields the scalar 
$\hat v\rfloor f$.

Already in the theory of equilibrium of, say, frames, beams, or 
plates, besides the balance of force, the balance of moment is required
to hold. A general distribution of forces at a rigid body can always 
be reduced to the action of a single {\it force} and a moment 
represented by a {\it couple} of equal but oppositely directed forces. 
This type of moment is independent of the reference point, i.e., it is 
described by a free (axial) vector (better: by a free antisymmetric tensor 
of 2nd rank). In this sense, a moment represented 
by a force couple has an independent status. Here, for the first 
time, besides the force, the moment (of a couple) steps in as 
new and `irreducible' quantity. The analogous happened in continuum 
mechanics: The force led to the concept of stress, and already at the 
end of last century, Voigt introduced additionally an independent 
couple stress (see [11]).

In modern continuum mechanics, when the description of crystalline solids, 
liquid crystals, molecular fluids, or layered media is at stake, the 
classical body of continuum mechanics is insufficient, which only can, if 
deformed, put on strains and nothing more. Continua with 
directors or some other additional structure are introduced. Departing 
mainly from Cosserat continua [4], this kind of continuum mechanics 
with microstructure, with directors, or micromorphic continuum mechanics,
or polar field theories,  
as it is variously called\footnote{There is quite an extended 
literature available on that subject. We found the following sources 
very useful: Capriz [2], Ericksen [8], Eringen and Kafadar [9], 
Ferrarese [10], Jaunzemis [19], Kr\"oner [21], Maugin and 
Eringen [23], Mindlin [27], and Truesdell and Toupin [33].}, has new 
degrees of freedom for deformation and, accordingly, new types of forces 
stressing such a continuum. 

This is how the concept of a double force (or of a force dipole) 
arose, adding to the couples (of forces) mentioned above, double 
forces without moments. In Fig.4, for the special case of 2 dimensions, 
we depicted a double force or, after the limiting procedure is conceived
to be done, a hyperforce, respectively. A hyperforce is represented 
by a 2nd rank tensor of type $(1,1)$, the upper (contravariant) index 
indicating the direction of the `lever arm', the lower (covariant) index 
marking the component of the force. In 3 dimensions, we have 9 different 
components of a hyperforce.

{}From the hyperforce $h^a{}_b$ we can covariantly split off the 
dilational hyperforce $h^c{}_c$. Left over is the deviatoric part 
$h^a{}_b-(1/3)\delta^a_b\,h^c{}_c$ which, if an underlying metric 
is available, can be split further into symmetric and antisymmetric 
parts. This decomposition 
$$h^{ab} = \underbrace{h^{[ab]}} 
_{\rm spin \> moment}
 +{1\over{3}}g^{ab} \underbrace{h^{c}{}_c}
_{\rm intrinsic \> dilational\> hyperforce}
+\underbrace{{\buildrel\frown\over{h\quer}}{}^{ab}}
_{\rm intrinsic \> shear \> hyperforce}.\eqno(\z)$$
is irreducible under the rotation group $SO(3)$. Therefore it is 
appropriate to attribute suitable names to each piece of the 
hyperforce. 

\begin{figure}[htb]
  \epsfbox[-10 0 500 270]{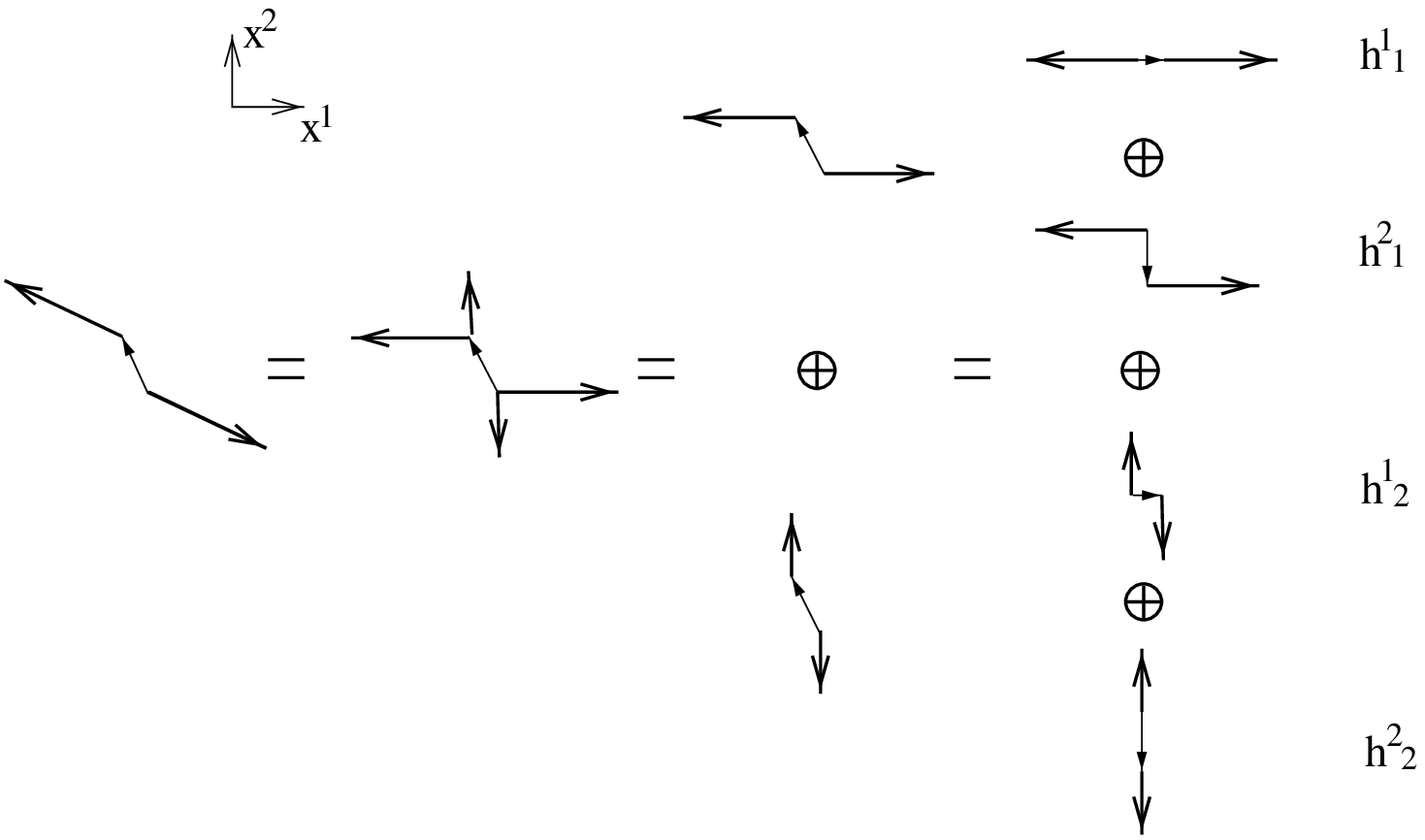} 
\label{decomp2d}
\end{figure}
\noindent{\bf Fig.4.} Hyperforce and its components in 2 dimensions: 
The Cartesian components of a double force or, in the limit, of an 
intrinsic hyperforce $h^a{}_b=\lim\Delta x^a\tilde f_b$. Note that 
$h^1{}_2$ and $h^2{}_1$ describe double forces with moment and 
$h^1{}_1$ and $h^2{}_2$ those without moment.

These considerations should be sufficient for supporting our 
thesis 1. Let us turn then to thesis 2.
\bigskip\goodbreak

\sectio{\bf Stress and Hyperstress as (n-1)-Forms}

\medskip
Stress is a concept which assigns, at some point $P$ of a continuum, 
to an arbitrary area element a corresponding force. In Fig.5 we 
illustrated this for a 3-dimensional continuum. We have to imagine 
that  the force is smeared over the area element.  Thus the stress 
caused   by   the   force   carries   the   dimension   of    {\it 
force/(length)}$^2$. 

\vspace*{2cm}

\begin{figure}[htb]
  \epsfbox[-10 0 500 270]{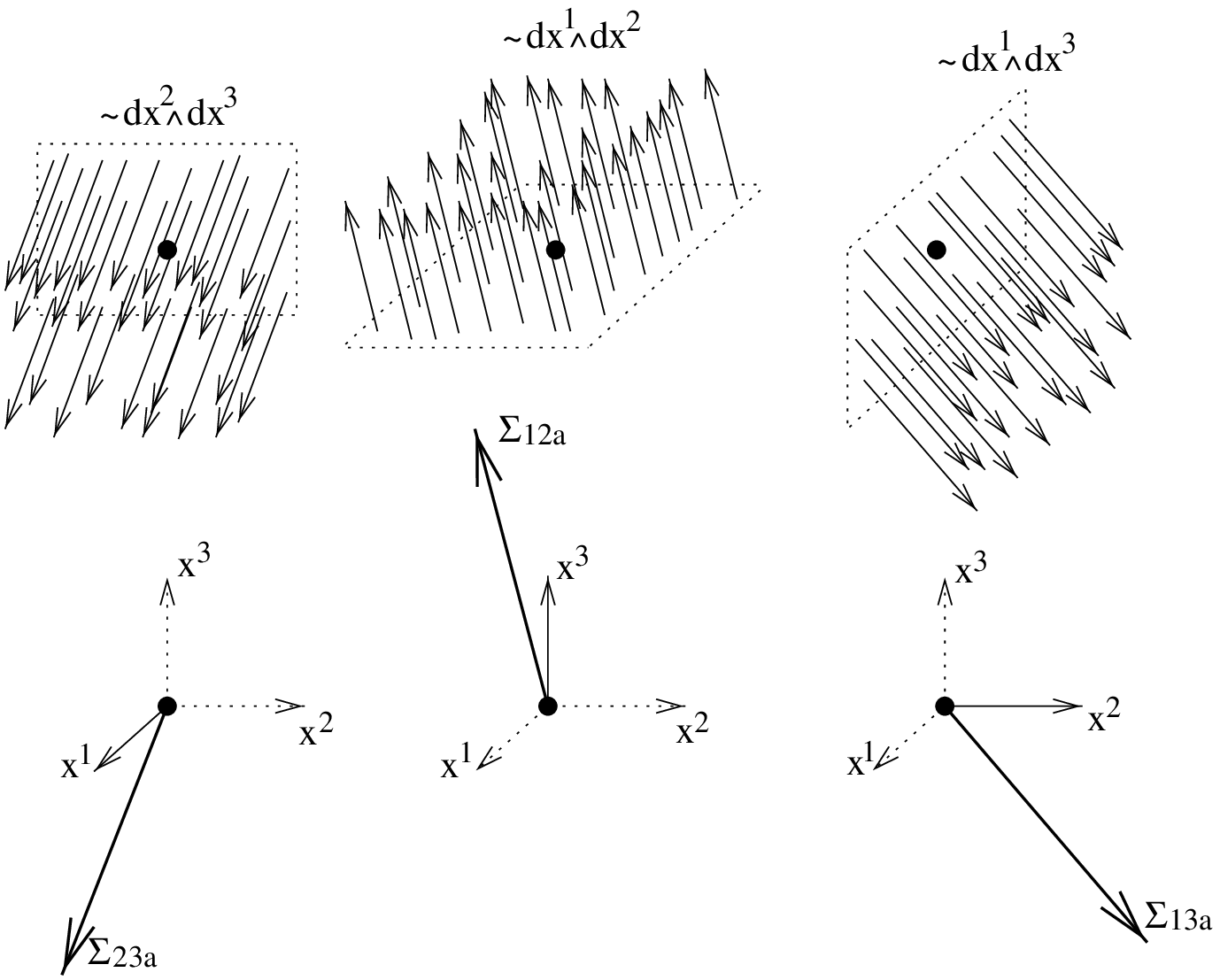} 
\label{stress}
\end{figure}
\noindent{\bf Fig.5.} Stress in a 3-dimensional continuum: The stress 
$\Sigma_a$ becomes a covector-valued 2-form. We use (holonomic) Cartesian 
coordinates $x^1,\,x^2,\,x^3$ and decompose $\Sigma_a$ according to 
$\Sigma_a=(1/2)\Sigma_{i_1i_2\,a}\,dx^{i_1}\wedge dx^{i_2}$. Under the 
state of stress $\Sigma_a$, a 2-dimensional area element 
$dx^{i_1}\wedge dx^{i_2}$, localized at a point P, responds with 
a force distribution smeared over this area element.

More specifically, using first the language of tensor 
analysis, at any point P, an area element 
$dA_a$ is linearly related to the force 
$dt^b$ distributed over $dA_a$ according to 
$$dt_b=\Sigma^a{}_b\;dA_a\,.\eqno(\z)$$
In this formalism, stress $\Sigma^a{}_b$ is a 2nd 
rank tensor of type (1,1). It represents the force that 
is transmitted through the area element $dA_a$ from its 
`minus' side to its `plus' side. 

Turning to the calculus of exterior differential forms, stress 
maps the area element $\delta{\vec u}\wedge\delta{\vec v}$ into 
a covector, see [15],
$$\delta t=\Sigma(\delta{\vec u},\delta{\vec v})\,.\eqno(\z)$$
Therefore $\Sigma$ is a covector-valued 2-form and can be 
decomposed as follows:
$$\delta t=\delta t_c\,\vartheta^c=\Sigma_c(\delta{\vec u},
\delta{\vec v})\,\vartheta^c\,\eqno(\z)$$
or
$$\delta t_c={1\over 2}\,\Sigma_{ab\,c}\,\vta^a\wedge\vta^b
\qquad{\rm with}\qquad\Sigma_{(ab)\,c}\equiv 0\,.\eqno(\z)$$ 
A collection of the most important notions in exterior calculus is 
given in the Appendix.

In 3 dimensions, $\eta_a$ of the $\eta$-basis (see Appendix) 
is a 2-form. Accordingly, we can 
decompose $\Sigma_b$ also with respect to $\eta_a$:
$$\Sigma_b=\Sigma^a{}_b\,\eta_a\,.\eqno(\z)$$
A comparison with (3.3) and (3.4) yields
$$\Sigma^a{}_b={1\over 2}\,\eta^{acd}\,\Sigma_{cd\,b}\,.\eqno(\z)$$
We recognize that the stress in either representation has 9 
independent components. 
Therefore the `engineering' stress in 3 dimensions is represented 
by a covector-valued 2-form 
$\Sigma_c=(1/2)\Sigma_{ab\,c}\,dx^a\wedge dx^b$ or, equivalently, 
by a 2nd rank tensor of type (1,1) $\Sigma^a{}_c$.

\begin{figure}[htb]
  \epsfbox[-10 0 500 240]{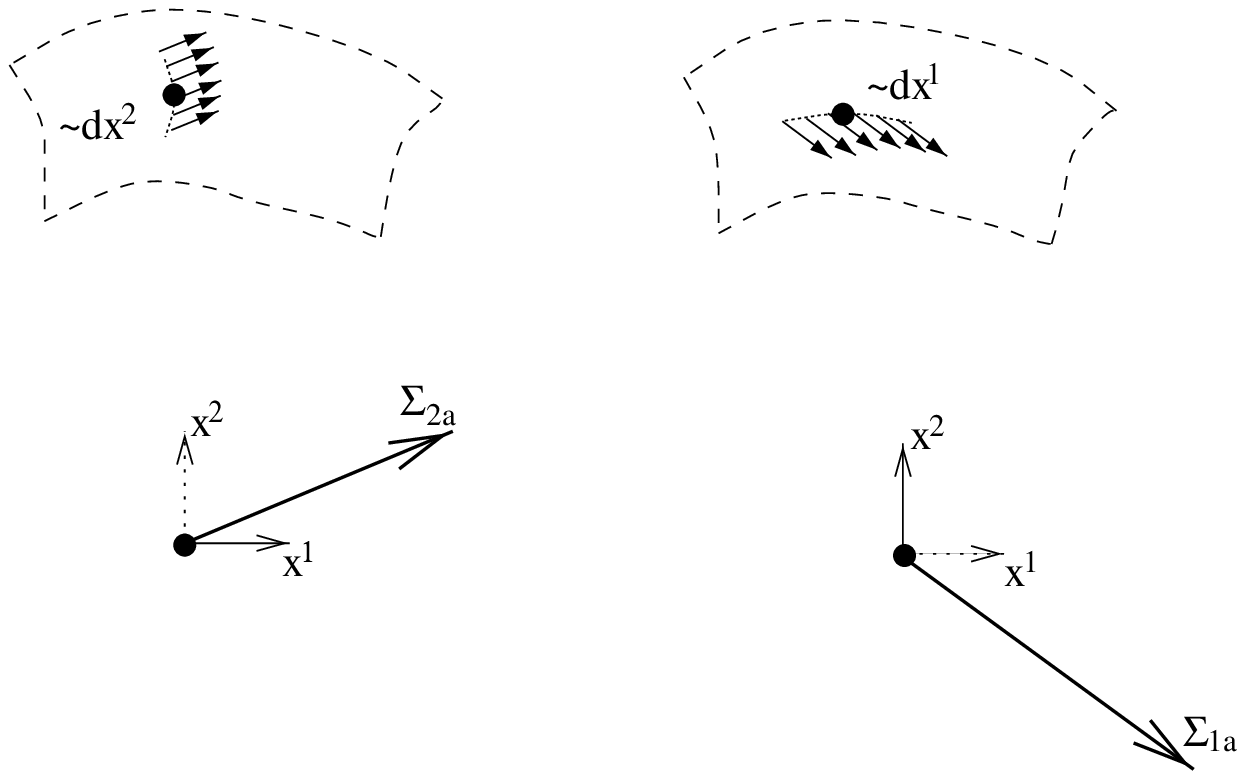} 
\label{stress2d}
\end{figure}
\noindent{\bf Fig.6.} Stress in a 2-dimensional shell: Stress turns 
into a covector valued 1-form $\Sigma_a=\Sigma_{i_1 a}\,dx^i$. 
The stress induces a force in the line element $dx^i$ that cuts 
the shell at a point P.

For the purpose of generalization it is useful to look for stress 
in the 2 dimensions of a shell (Fig.6) and in the 1 dimension of a 
wire (Fig.7), for example. In Fig.6 we cut a shell 
along a 1-dimensional line. 
Stress assigns then a 2-dimensional force to the 1-dimensional 
line. Analogously, in the case of the wire in Fig.7, a 
1-dimensional force is assigned to the 0-dimensional cross section 
of the wire. In other words, for up to 3 dimensions, stress is a 
covector-valued $(n-1)$-form. The `area' element in question is 
always a hypersurface in the corresponding continuum, which 
separates two parts of the continuum from each other. The stress 
is a measure of the force transmitted through that area element 
from one part of the continuum to the other part. The 
force remains a covector in any dimension.

\begin{figure}[htb]
  \epsfbox[-10 0 500 120]{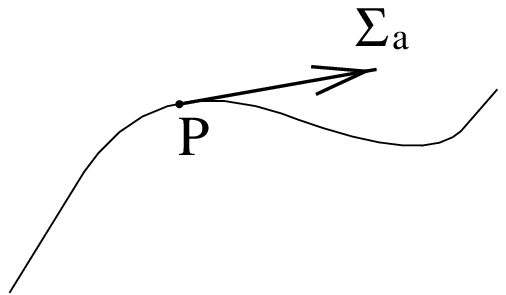} 
\label{stress1d}
\end{figure}
\noindent{\bf Fig.7.} Stress in a 1-dimensional wire: Stress 
degenerates to a covector valued 0-form. Here the force at P 
coincides with the stress.

The situation is quite similar in spacetime physics. In the special 
case of Maxwell's 
theory, we saw already that the 4-dimensional generalization of the 
Maxwell stress is given by the momentum current of the free 
electromagnetic field. A momentum current is a measure for the flow 
of momentum in a 
3-dimensional volume element. Because momentum and force are related
by differentiation, the momentum, like force, is a covector-valued 
form. Consequently we introduce the momentum current -- as some 
kind of 4-dimensional stress in spacetime -- as a covector-valued 
3-form according to $(\alpha , \beta...=0,1,2,3)$ 
$$\Sigma_\delta ={1\over {3!}}\Sigma_{\alpha \beta \gamma\, \delta}\,
\vartheta^\a\wedge\vartheta^{\b}\wedge\vartheta^{\c}
=\Sigma^{\alpha}{}_{\delta}\,\eta_{\a}\,,\eqno(\z)$$
or
$$\Sigma^{\alpha}{}_{\delta}={1\over{3!}}\eta^{\mu \nu \sigma \a}
\,\Sigma_{\mu \nu \sigma\,\delta}\,.\eqno(\z)$$
How one arrives at the 4-dimensional concept of a momentum current 
in field theory is very transparently discussed by Dodson and Poston [5] 
from a geometrical point of view.

So far, we have supported the stress part of our thesis 2. Let us 
now turn to the hyperstress part. The classical body of elasticity 
theory in 3-dimension or the Riemannian spacetime continuum 
of Einstein's theory (in 4 dimensions) cannot carry hyperstress, 
that is, hyperstress vanishes under these 
circumstances. Strictly speaking, this remark needs 
qualification in the case of spacetime. In discussing the Dirac 
field and its currents, we will come back to it in Sec.5. 
\vspace*{2cm}

\begin{figure}[htb]
  \epsfbox[-10 0 500 240]{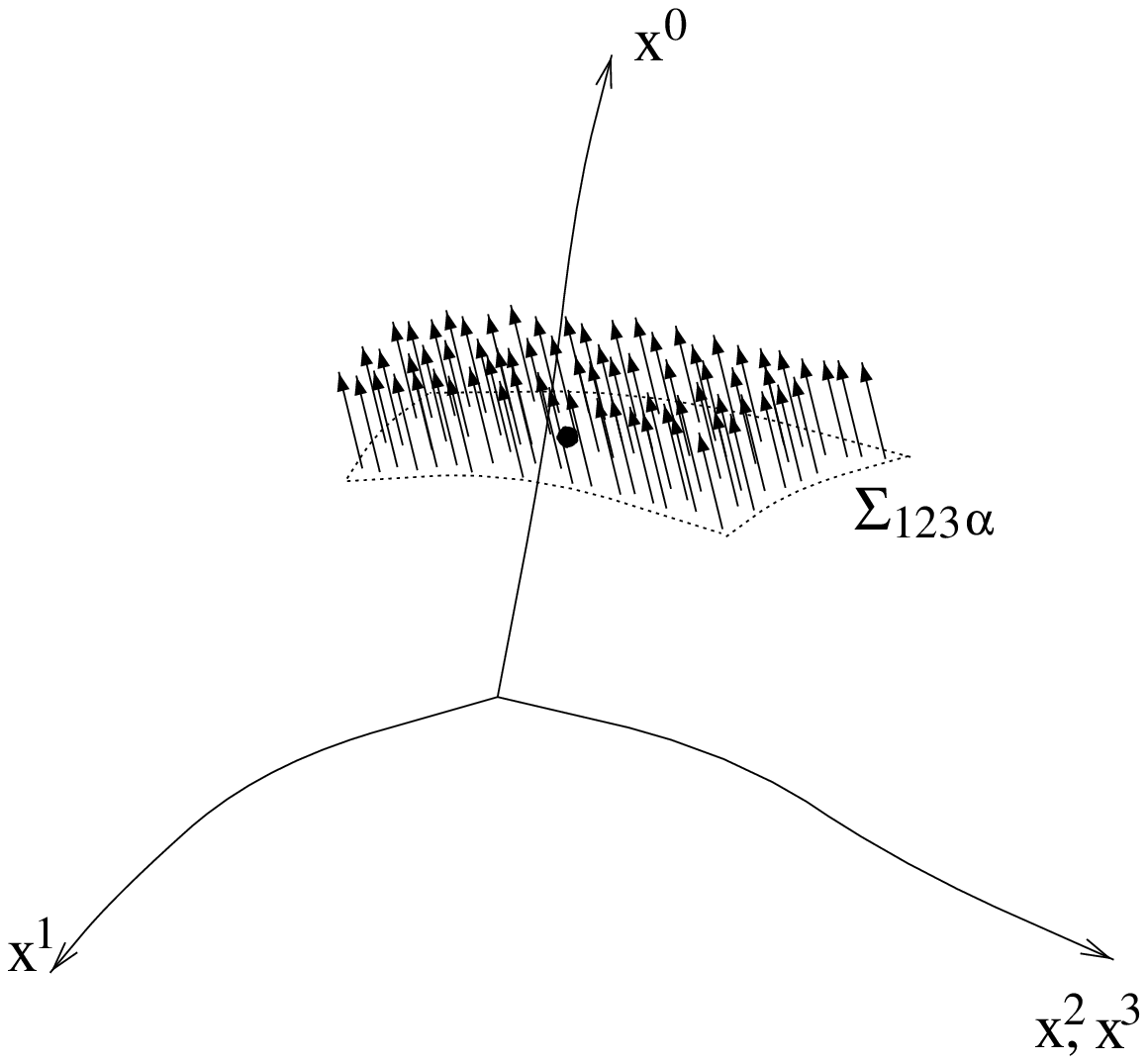} 
\label{stress4d}
\end{figure}
\noindent{\bf Fig.8.} Stress as momentum current in 4-dimensional 
spacetime: The 3-dimensional (spacelike) hypersurface element 
carries a distribution of momenta.

Accordingly, in order that a continuum can carry hyperstress, we 
have to assume that it has additional deformational degrees of 
freedom. Consider Mindlin's 3-dimensional theory of 
`Micro-structure in Linear Elasticity' [27] as one of the most 
illuminating models in this context. His model continuum carries 
both stress and hyperstress and its 
deformation is described by three different types of deformational 
measure: a macrostrain (6 components), a relative deformation 
(the difference between the macro-displacement-gradient and the 
microdeformation, 9 components), and a micro-deformation gradient 
(27 components). 

Consequently, if one wanted to generalize these concepts in such a 
way that they also apply to the 4-dimensional spacetime continuum, 
one would have, in 4 dimensions, $(10+16+64)$ deformational measures, 
namely -- that is our `guess' -- metric plus coframe plus connection.
That the micro-deformation gradient really relates to the connection 
1-form of a non-Riemannian spacetime, can be seen as follows: A frame, 
which is the analog of a director field of microstructured continuum, 
if parallelly transported by means of the underlying connection, see 
(A.1), deforms linearly, it {\it rotates}, {\it dilates}, and 
undergoes a {\it shear}. This 
deformation is coded into the connection as projected to the frame, 
that is, into the frame components of the connection of spacetime. 
Therefore, an independent $GL(n,R)$-valued connection 1-form 
$\Gamma_\alpha{}^\beta=\Gamma_{\gamma\alpha}{}^\beta\,\vartheta^\gamma$ 
will do the job; here we switched immediately to $n$ dimensions. 
Mindlin's theory could be rewritten in terms of these geometrical 
notions.

Therefore we will assume that spacetime represents a continuum 
with {\it microstructure}. The idea is to attach to each point of 
an $n$-dimensional 
continuum $n$ linearly independent vectorial `directors'. Besides 
the displacement field, the motion and, in particular, the 
deformation of the directors, coded into the linear connection 
of the material manifold, specify the state of the continuum. 
The response of the continuum to the deformation of the directors, 
if conceived as an elastic body, consists of {\it hyperstress}.
\vspace*{3cm} 

\begin{figure}[htb]
  \epsfbox[-10 0 500 220]{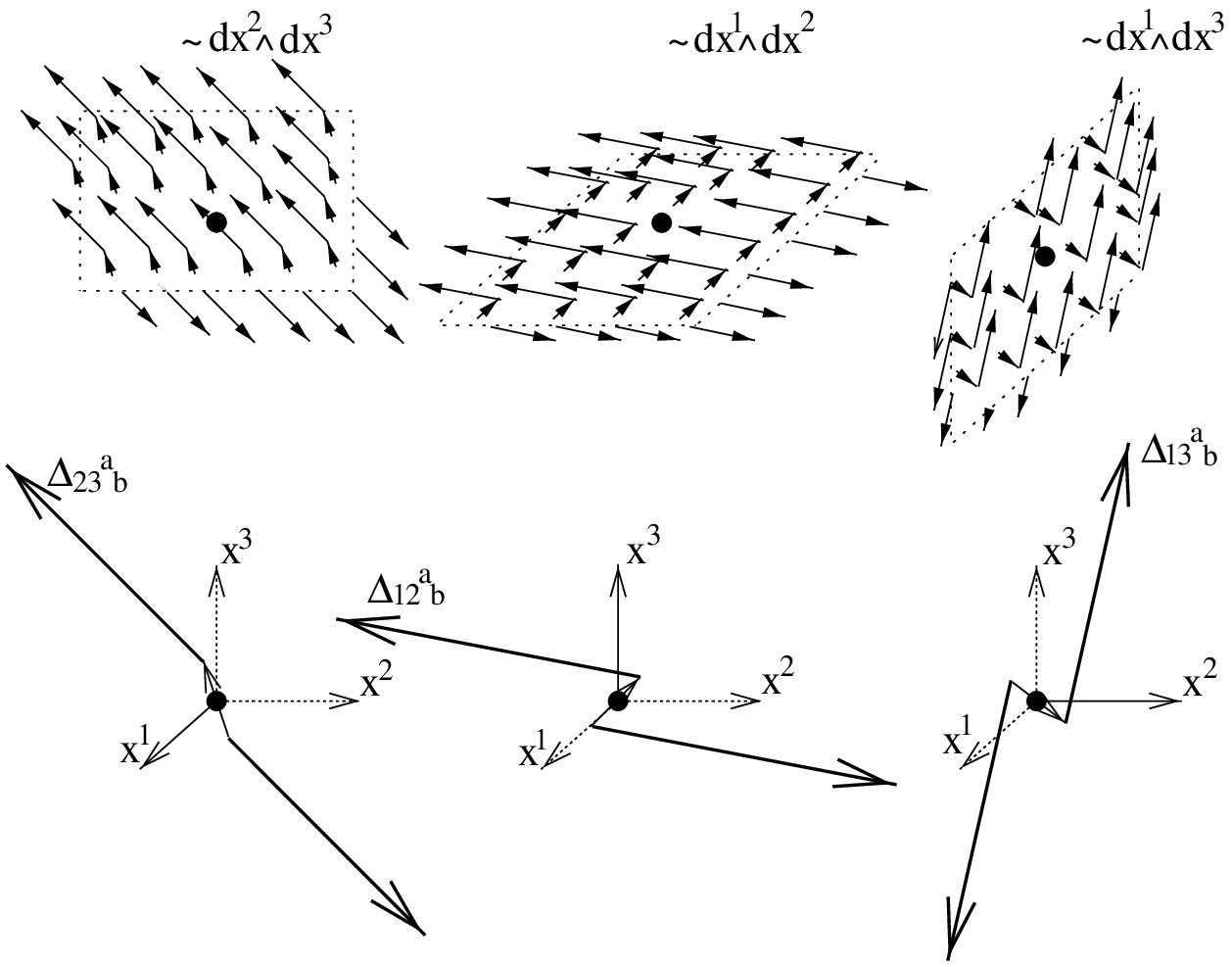} 
\label{hypstress}
\end{figure}
\noindent{\bf Fig.9.} Hyperstress in a 3-dimensional continuum: 
The (1,1)-tensor valued 2-form $\Delta^\a{}_\b=(1/2)
\Delta_{i_1i_2}{}^\a{}_\b\,dx^{i_1}\wedge dx^{i_2}$ induces a 
hyper\-force into the infini\-tesimal 
2-dimen\-sional area element $dx^{i_1}\wedge dx^{i_2}$, 
compare Fig.5 for the analogous case of stress.

In analogy to (3.1), compare Fig.9 and Fig.5, we introduce in 3 
dimensions the intrinsic hyperstress tensor 
$\Delta^{a\,b}{}_c$ according to
$$dh^b{}_c=\Delta^{a\,b}{}_c\,dA_a\eqno(\z)$$
and the corresponding hyperstress 2-form as 
$$ \Delta^b{}_c={1\over 2}\,\Delta_{de}{}^b{}_c\,\vartheta^d\wedge
\vartheta^e\,,\quad\qquad \Delta_{de}{}^b{}_c=
-\Delta_{ed}{}^b{}_c\,,\eqno(\z)$$
with
$$\Delta^{a\,b}{}_c={1\over 2}\,
\eta^{ade}\,\Delta_{de}{}^b{}_c\,.\eqno(\z)$$

In 4 dimensions, we find for the intrinsic hypermomentum current
$$ \Delta^\b{}_\c=\Delta^{\a\,\b}{}_\c\,\eta_\a\qquad
\quad{\rm or}\quad\qquad\Delta^{\a\,\b}{}_\c={1\over 3!}
\,\eta^{\mu\nu\sigma\a}\,
\Delta_{\mu\nu\sigma}{}^\b{}_\c\,,\eqno(\z)$$
the generalization to arbitrary $n$ dimensions being obvious. 
Incidentally, the orbital hypermomentum will be defined in (4.4).
Thus, in analogy to the momentum current $\Sigma_\c$, we 
arrived at the concept of a hypermomentum current 
$\Delta^\b{}_\c$. Obviously, the notion of hyperstress, 
as (1,1) tensor-valued $(n-1)$-form, is firmly anchored in modern 
versions of theories of continuua with microstructure.

\sectio{\bf Hypermomentum Current}
\medskip

The intrinsic hypermomentum current (3.12), in analogy to (2.1), 
can be decomposed into 3 different pieces:
$$\Delta^{\alpha\beta}=\underbrace{\tau^{\alpha\beta}}_{\rm spin\> 
current}+\>\>{1\over 4}\,g^{\alpha\beta}\,
\underbrace{\Delta^\c{}_\c}_{\rm 
intrinsic\> dilation\> current} +
\underbrace{{\buildrel\frown\over{\Delta\quer}}{}^{\alpha\beta}}_
{\rm intrinsic\> shear\> current}\,.\eqno(\z)$$ 

In the Poincar\'e gauge theory of gravity, the dynamical spin 
$\tau^{\alpha\beta}:= \Delta^{[\alpha\beta]}$ is a source term 
which is, besides the momentum current $\Sigma_{\alpha}$, 
of crucial importance (see [22, 24, 26, 32]). Moreover, the dilation current 
$\Delta:=\Delta^\c{}_\c$ is an essential ingredient of conformal 
models of gravity based on a Weyl geometry (see [17]) and is familiar 
from canonical field theory. Namely, we know from special-relativistic 
Lagrangian field theory the scale or dilation current of, say, a 
scalar field $\psi(x)$ 
with scale dimension $S$. Let a local conformal change of the metric 
be $g_{ij}\rightarrow \phi^L(x)\,g_{ij}$. Then the field transforms 
as $\psi(x) \rightarrow \tilde \psi(\tilde x)$ = $\phi   
^S\; \psi\bigl( \phi ^{L/2}\, x\bigr) $ 
under scale transformations and the dilation current, in 
Cartesian coordinates, reads     
$$ \Upsilon:= {\partial {L} \over \partial D \psi} 
S \psi + x^\c\Sigma _\c=\Delta+ x^\c \Sigma _\c\> .\eqa $$ 
The total current $\Upsilon$ consists of an intrinsic part $\Delta$ 
and an orbital part, a product of the radius `vector' and the 
canonical momentum current $\Sigma_\c$ of the scalar matter 
field. Provided the action is  
scale invariant, the dilation current is conserved. Its charge  
generates the dilation group. In the gauge approach to the Weyl  
group, this current, which, of course, reflects the behavior   
of the matter field under {\it scale} transformations, couples to the  
Weyl covector.

Only the last piece on the right hand side of (4.1), the intrinsic shear 
$${\buildrel\frown\over{\Delta\quer}}{}^{\alpha\beta} := 
\Delta^{(\alpha\beta )} - 
{1\over 4}g^{\alpha\beta}\,\Delta\,,   \eqno(\z)$$ 
is more remote from 
{\it direct} physical experience. In the manifield approach of 
Ne'eman et al.\ [28, 29, 30, 17] it manifests 
itself indirectly in the occurrence of states within the 
infinite--dimensional representations of the $SL(4,R)$ which 
are lying along Regge trajectories (see [7]). In eq.(6.9) and the 
subsequent discussion we will present some further evidence 
for the existence of a shear current.

Besides the intrinsic pieces of the hypermomentum current, quite 
analogous to what happens in the case of the angular momentum 
current and the dilation current in (4.2), there must 
exist corresponding orbital pieces. Therefore we postulate (in 
Cartesian coordinates) for the total hypermomentum current: 
$$\Upsilon^{\alpha}{}_{\beta}:{\buildrel {*}\over {=}}
\,\Delta^{\alpha}{}_{\beta}+x^{\alpha}\,\Sigma_{\beta}\,.\eqno(\z)$$
Its trace is the total dilation current,
$$\Upsilon:=\Upsilon^{\gamma}{}_{\gamma}\;{\buildrel {*}\over {=}}\;
\,\Delta+x^{\gamma}\,\Sigma_{\gamma}\,, \eqno(\z)$$
its antisymmetric piece the total angular momentum current,
$$\Upsilon^{[\alpha \beta]}\,{\buildrel {*}\over {=}}\,
\tau^{\alpha \beta}+x^{[\alpha}\,\Sigma^{\beta ]}\,.\eqno(\z)$$ 
The new piece is the shear current,
$${\buildrel\frown\over{\Upsilon\quer}}{}^{\alpha\beta} 
:=\Upsilon^{(\alpha \beta)}-
{1\over 4}g^{\alpha \beta}\,\Upsilon^{\gamma}{}_{\gamma}=
{\buildrel\frown\over{\Delta\quer}}{}^{\alpha\beta} 
+\{x^{(\alpha}\,\Sigma^{\beta )}-{1\over 4}
g^{\alpha \beta}\,x^\gamma\,\Sigma_\gamma\}\,.\eqno(\z)$$
We display all these currents in a small table:
$$\vbox{\offinterlineskip
\hrule
\halign{&\vrule#&\strut\quad\hfil#\quad\hfil\cr
height2pt&\omit&&\omit&&\omit&\cr
& {\bf total current} && {\bf intrinsic piece} && {\bf orbital piece} &\cr
height2pt&\omit&&\omit&&\omit&\cr
\noalign{\hrule}
height2pt&\omit&&\omit&&\omit&\cr
& hypermomentum $\Upsilon^{\alpha}{}_\beta$ && 
$\Delta^{\alpha}{}_\beta$ && $x^{\alpha}\,\Sigma_{\beta}$ &\cr
height2pt&\omit&&\omit&&\omit&\cr
\noalign{\hrule}
height2pt&\omit&&\omit&&\omit&\cr
& dilation $\Upsilon$ && $\Delta$ && $x^\c\,\Sigma_\c
$ &\cr & angular momentum  $\Upsilon^{[\alpha \beta ]}$  && 
$\tau^{\alpha \beta}$ && $x^{[\alpha }\,\Sigma^{\beta ]}$ &\cr
& shear ${\buildrel\frown\over{\Upsilon\quer}}{}^{\alpha\beta} 
$ && ${\buildrel\frown\over{\Delta\quer}}{}^{\alpha\beta} 
$ && $x^{(\alpha }\,\Sigma^{\beta )}-{1\over 4}g^
{\alpha \beta}\,x^{\gamma}\,\Sigma_{\gamma}$ &\cr 
height2pt&\omit&&\omit&&\omit&\cr}
\hrule}$$

The considerations of this and the last section establish stress and 
hyperstress as fundamental concepts in field theory, as we claimed 
in thesis 2. Nevertheless, for illustrative purposes we will turn to 
the Dirac field and will first show that its spin current 
represents one piece of the hypermomentum current. 
\bigskip
\bigskip\goodbreak

\sectio{\bf Dirac Field and its Momentum and Spin Currents}
\medskip

As we saw in Sec.1, the electromagnetic 
field historically was the first physical system which led to the 
notion of a 4-dimensional momentum current (see also the review 
[13]). In vacuum, the momentum current obeys the conservation laws
$$D\,{\buildrel{\rm max}\over{\sigma_\a}}=0\,,\qquad\qquad
\vta_{[\a}\wedge{\buildrel{\rm max}\over{\sigma_{\b]}}}=0\,,\eqa$$ 
and is, in addition, tracefree:
$$\vta^\a\wedge{\buildrel{\rm max}\over{\sigma_\a}}=0\,.\eqa$$
Evidently, ${\buildrel{\rm max}\over{\sigma_\a}}$ is a well-%
behaved stress which, apart from (5.2), is reminiscent of the 
Euler-Cauchy stress of continuum mechanics.

In 1915, Hilbert and Einstein quite generally postulated the existence 
of an (energy-)momentum current $\sigma_\a$ linked to any form of 
matter. This current, as it was assumed in analogy to Maxwell's field, 
should be symmetric, $\vta_{[\a}\wedge\sigma_{\b]}=0$, and conserved, 
$D\sigma_\a=0$. Ordinary macroscopic matter can be described 
hydrodynamically, and, under these premises, a symmetric and conserved 
energy-momentum current $\sigma_\a$ is appropriate and successful in 
subsuming the inertial properties of matter. 

In the late twenties, however, the Dirac field was discovered. In 
this context, the symmetry and conservation of the momentum current had to 
be reconsidered. Let us describe the Dirac field in exterior forms, 
see the Appendix and [14]. The Dirac matrices, referred to 
an orthonormal frame, obey the anti-commutation rule 
$$\gamma_\alpha\gamma_\beta+\gamma_\beta\gamma_\alpha=2o_{\alpha\beta}
  \bbbone\,.\eqno(\z)$$
The six matrices
$$\sigma^{\alpha \beta}:= {i\over{4}}\gamma^{[\alpha}\gamma^{\beta]}
\eqno(\z)$$
are the infinitesimal generators of the Lorentz group and fulfill the
commutation relation
$$[\sigma_{\alpha \beta},\gamma_{\c}] = {1\over{2}}
\eta_{\alpha \beta \gamma\delta}\,\gamma_5\gamma^{\delta}
\,,\qquad\gamma_5:=\gamma_0\gamma_1\gamma_2\gamma_3 \,\,.\eqno(\z)$$
We recall, that the 16 elements 
$\{\bbbone,\gamma_\alpha,\sigma_{\alpha\beta},\gamma_5,
\gamma_5\gamma_\alpha\}$ of $4\times 4$ matrices form a basis of a
 Clifford algebra. The constant $\gamma_\alpha$ matrices can be
converted into Clifford algebra-valued $1$- or $3$-forms, respectively:
$$\gamma:=\gamma_\alpha\,\vartheta^\alpha\,,\qquad \hodge\gamma
=\gamma_\alpha\,\eta^\alpha\,.\eqno(\z)$$
The exterior covariant derivative D acts on $\Psi$ as
$$D\Psi=d\Psi+{i\over 4}\Gamma^{\alpha\beta}\wedge\sigma_{\alpha\beta}\Psi
  \quad\Longrightarrow\quad
 \overline{D\Psi}=d\aPsi-{i\over 4}\Gamma^{\alpha\beta}\wedge\aPsi
  \sigma_{\alpha\beta}\,.\eqno(\z)$$

Let us first consider quite generally an isolated matter system 
in SR (special relativity) 
with a general Lagrangian 4-form 
$L = L( o_{\alpha\beta}, \vartheta^\alpha,\Psi, D\Psi)$. We stipulate 
that the Euler-Lagrange equation $\delta 
L/\delta\Psi = 0$ for the matter field $\Psi$ is fulfilled. Then, 
global or {\it rigid} Poincar\'e invariance yields, due to 
the Noether theorem, the conservation laws for energy-momentum 
$$D\Sigma_\alpha=0 \eqno(\z)$$
and angular momentum
$$ D\tau_{\alpha\beta}+\vartheta_{[\alpha}\wedge
\Sigma_{\beta]} \,\,{\buildrel {*}\over {=}} 
\,\,D\bigl(\tau_{\alpha\beta}+ x_{[\alpha}\wedge
\Sigma_{\beta]}\bigr)=0\,,\eqno(\z)$$
with the {\it canonical} (or Noether) energy-momentum current
$$\Sigma_{\alpha}:= 
 e_{\alpha}\rfloor L -  (e_{\alpha}\rfloor  
D\Psi)\wedge{{\partial L}\over{\partial (D\Psi)}} -  
(e_{\alpha}\rfloor\Psi)\wedge{{\partial L}\over{\partial\Psi}}
\eqno(\z)$$  
and the canonical (or Noether) spin current
$$\tau_{\alpha \beta}:=  \rho(\Lambda
_{\alpha \beta})\,\Psi\wedge{{\partial L}\over{\partial 
(D\Psi)}} \,,\eqno(\z)$$  
respectively. The star indicates that the equality is valid only 
in a Cartesian coordinate system, and $\rho(\Lambda_{\alpha \beta})$ 
denotes a suitable representation of the Lorentz group.                   

We assume that the currents fall off sufficiently fast towards 
space-like infinity. Then, by integrating the currents over a
spacelike hyperplane $H_t$, we find the conserved `charges'
$$P_{\alpha}\asteq \int_{H_t}\Sigma_{\alpha}\,,
  \qquad {\rm total \> energy\!-\!momentum}\,,\eqa$$
and
$$J_{\alpha \beta}\asteq \int_{H_t}(\tau_{\alpha \beta}
+x_{[\alpha} \Sigma_{\beta ]})\,,\qquad {\rm total \> angular \> momentum}
\,, \eqno(\z) $$
respectively.

Explicitly, the Dirac Lagrangian is given by the $4$-form
$$L_D=L(\vartheta^\alpha,\Psi,D\Psi)=
  {i\over 2}\left\{\aPsi\,\hodge\gamma\wedge D\Psi
  +\overline{D\Psi}\wedge\hodge\gamma\,\Psi\right\}+\hodge m\,\aPsi\Psi\,,
  \eqa$$
for which $L_D=\overline{L}_D=L_D^\dagger$, as required. The coframe
$\vartheta^\alpha$ necessarily occurs in the Dirac Lagrangian, even in SR.
For the mass term, we use the short-hand notation $\hodge m=m\eta$. The
hermiticity of the Lagrangian in (5.14) leads to a charge current
which admits the usual probability interpretation.
The Dirac equation is obtained by varying $L_D$
with respect to $\aPsi$:
$$i\,\hodge\gamma\wedge D\Psi+\hodge m\,\Psi-{i\over 2}(D\hodge\gamma)\Psi
  =0\,.\eqa$$
In flat Minkowski spacetime, the expression $D\,\hodge\gamma$ vanishes 
because of $D\eta_\a=0$. Then the Dirac equation reduces to
$$\hodge\gamma\wedge D\Psi=i\,\hodge m\,\Psi\,.\eqa$$

The canonical currents (5.10) and (5.11) read, respectively:
$$\Sigma_\alpha={i\over 2}\left\{\aPsi\,\hodge\gamma\wedge D_\alpha\Psi
  -\overline{D_\alpha\Psi}\wedge\hodge\gamma\Psi\right\}\,,\qquad
  D_\alpha:=e_\alpha\rfloor D\,,\eqa$$
$$\tau_{\alpha\beta}=\tau_{\alpha\beta\gamma}\,\eta^\gamma
={1\over 8}\aPsi\left(\hodge\gamma\sigma_{\alpha\beta}+
  \sigma_{\alpha\beta}\hodge\gamma\right)\Psi
  ={i\over 4}\,\aPsi\gamma_{[\alpha}\gamma_\beta\gamma_{\gamma]}\Psi\,
  \eta^\gamma\,,\eqa$$
This implies that the components
$\tau_{\alpha\beta\c}=\tau_{[\alpha\beta\c]}$ of the spin current are
totally antisymmetric. Therefore again, as in the Maxwell case,
$\vartheta_{(\alpha}\wedge\Sigma_{\beta)}$ represents the symmetric
momentum current which is coupled to the metric of spacetime.
However, because of (5.9) and of $D\tau_{\a\b}\neq 0$, 
there is a genuine antisymmetric part $\vta_{[\a }\wedge\Sigma_{\b]}$ of
the momentum current which is induced by the Dirac spin.
Consequently, with the Dirac field we found a physical system 
which carries, besides the momentum current, also a non-vanishing spin 
current. And, due to the spin current, the momentum current loses 
its symmetry, as displayed in the angular momentum law (5.9). 

The canonical currents
$(\Sigma_{\alpha},\tau_{\alpha \beta})$ are 
only determined up to an exact form. This enables us to introduce 
superpotentials and to perform relocalization transformations that 
leave the Noether identities as well as the conserved 
charges invariant and lead to a symmetric momentum current, see 
[14, 20]. Nevertheless, the canonical currents $(\Sigma_{\alpha},
\tau_{\alpha \beta})$ are those of primary importance and the 
symmetric momentum current is only a truncated auxiliary quantity 
which no longer carries the momentum contribution induced by spin. 

With the spin current (5.18) of a Dirac field, we found an example for  
the first term on the right hand side of the hypermomentum current 
in (4.1) and the corresponding entry in our table. By the same token, we 
found, as earlier in continuum mechanics, an asymmetric momentum current, 
namely $\Sigma_\a$ of (5.17). For anybody with some background 
knowledge in continuum mechanics, there can be no doubt that the existence 
of the inertial Dirac currents $(\Sigma_{\alpha},\tau_{\alpha \beta})$ 
indicate that spacetime should carry, besides the Riemannian metric 
$g$, a microstructure of the type of Mindlin's continuum, at least 
if the Dirac field has to be considered as a generic matter field.  

Still, because of the possibility of constructing a symmetric 
$\sigma_\a$ for the Dirac field out of the canonical current 
$\Sigma_{\alpha}$,
the corresponding 2nd rank energy-momentum current can be coupled to 
the metric of a Riemannian spacetime, as it happens in GR, indeed. 
However, only if we postulate a `microstructure' of spacetime, as is 
done in the Poincar\'e gauge theory with its Riemann-Cartan spacetime, a 
coupling of the spin $\tau_{\a\b}$ to the 
{\it contortion} of spacetime can take place, as it would seem 
desirable for a piece of the 4-dimensional `hyperstress'.

Up to now we found evidence in favor of the existence of the spin and 
the intrinsic dilation current, which both emerge in (4.1). Subsequent 
to (4.3), we made some remarks on the intrinsic {\it shear} current. 
We feel, however, that some plausibility considerations could help to 
uncover how natural the concept of the shear current really is. In 
this section, we only used the {\it antisymmetric} part 
$\int x^{[\alpha}\,\Sigma^{\beta ]}$ of the dipole moment 
$\int x^{\alpha}\,\Sigma_{\beta}$ of the momentum current, see (5.13). 
Let us try to get ideas about the corresponding symmetric part.
\bigskip\goodbreak

\sectio{\bf Quadrupole Excitations, Signorini's Mean Stresses}
\medskip

There is a close connection between {\it quadrupole excitation} and the 
shear current. Spatially extended material particles -- whether molecules,
nuclei, or hadrons -- display rotational and vibrational
excitations. The spectra have been studied in molecular, nuclear,
and hadronic physics, and it is often observed that they form
bands in which the spin levels occur with intervals of two angular
momentum quanta $\Delta j = 2$. This feature points to a quadrupolar
excitation mechanism, as it takes place in the pulsations of a deformed 
(sometimes cigar-shaped) nucleus or in the Regge trajectories of 
the quark-antiquark or three-quark structures of the hadrons. 
The expression $x^{[\alpha}\,\Sigma^{\beta]}$, by contrast, 
displays dipolar features of orbital angular momentum and, by
extension, of the spin itself; rotational excitation bands indeed
display intervals of $\Delta j=1$, characteristic of dipolar mechanisms.
These different states have been classified with the help of dynamical 
groups and spectrum generating algebras. 

Dothan et al.~[7] were the first to discuss the $SL(3,R)$ as a 
dynamical group for the classification of hadronic excitations. This was 
later on improved and extended by Ne'eman and \v Sija\v cki [28, 29, 30]. 
Following them, it is therefore natural to look for a classical 
realization of the {\it deformation}-currents based on the action
of quadrupoles; a current is a charge-flow and for the deformations 
one looks at the pulsation rate, i.e. the time-derivative
of the mass-quadrupole moment,
$${d\over dt}\int x^\alpha x^\beta\,\Sigma^{k\gamma}\,dA_k\, ,\qquad
(\rm with\>\Sigma^{\c}=\Sigma^{\kappa\c}\,\eta_{\kappa})\,.\eqno(\z)$$
We are using here tensor calculus in Cartesian coordinates within the 
framework of SR. Let us differentiate the integrand in (6.1) 
with respect to the {\it second} 
index of the energy-momentum tensor:
$$\partial_{\gamma}(x^\alpha x^\beta \Sigma^{k\gamma})=
x^\alpha \Sigma^{k\beta}+x^\beta\Sigma^{k\alpha}
+x^\alpha x^{\beta}\partial_{\gamma}\Sigma^{k\gamma}\,.\eqno(\z)$$
But energy-momentum conservation yields
$$\partial_{\gamma}\Sigma^{\gamma k}=0=\partial_{\gamma}\Sigma^{(\gamma k)}
+\partial_{\gamma}\Sigma^{[\gamma k]}=\partial_{\gamma}\Sigma^{(k\gamma)}
-\partial_{\gamma}\Sigma^{[k\gamma]}=\partial_{\gamma}\Sigma^{k\gamma}
-2\partial_{\gamma}\Sigma^{[k\gamma]} \,,\eqno(\z)$$
$$\Longrightarrow \qquad \partial_{\gamma}\Sigma^{k\gamma}
= 2\partial_{\gamma}\Sigma^{[k\gamma]} \,. \eqa$$
We substitute (6.4) in (6.2)
$$\partial_{\gamma}(x^\alpha x^{\beta} \Sigma^{k\gamma})=
2\{x^{(\alpha}\Sigma^{|k |\b)}+x^{\alpha}x^{\beta}
\partial_{\gamma}\Sigma^{[k\gamma]}\}\,. \eqno(\z)$$  

Eq.(6.5) can be simplified in the case of a Dirac field. As above, we 
introduce the spin tensor $\tau^{\a\b\c}$ by $\tau^{\alpha \beta}=
\tau^{\alpha \beta \gamma}\eta_{\gamma}$.
Differentiation of angular momentum law in SR yields
$$\partial_{\beta}\partial_{\gamma}\tau^{\c\a\b}+
\partial_{\beta}\Sigma^{[\alpha \beta ]}=0\, . \eqno(\z)$$
However, for a Dirac field we have
$$\tau^{\alpha \beta \gamma}=\tau^{[\alpha \beta \gamma ]}\qquad
{\Longrightarrow}\qquad \partial_{\beta}\Sigma^{[\alpha \beta ]}=0
\,.\eqa$$
Substitution of (6.7) into (6.5) leads to the decisive formula:
$$\partial_{\gamma}(x^{\alpha} x^{\beta}\Sigma^{k\gamma})=
 2x^{(\alpha}\Sigma^{|k|\beta )}\,. \eqno(\z)$$
We integrate this relation over a spacelike hypersurface.
If the Dirac field represents an insular matter distribution, we end 
up with 
$${d\over dt}\,\int d^3\!x\,\, x^\alpha x^\beta\Sigma^{ k0}=2\int 
d^3\!x\,\,x^{(\alpha}\Sigma^{|k|\b)}\,.\eqno(\z)$$
For a Dirac field, according to (6.9), the rate 
of change of the quadrupole moment can thus 
be expressed in terms of the symmetric part 
$x^{(\a}\Sigma^{|k|\beta)}$ of the orbital 
hypermomentum current. Consequently,
the quadrupolar excitations of Dirac particles suggest the 
existence of the shear current, at least of the orbital part.
But then, in analogy to the angular momentum and the dilation current, 
the conclusion is hard to circumvent that in general an intrinsic 
piece of the shear current should accompany the orbital piece.

On the right hand side of (6.9), if one concentrates only on the
spatial components of the expression, we have a typical integral 
defining {\it Signorini's mean stresses}, see page 150 of Truedell's 
textbook [34]. In fact, it was Signorini's theory of mean stresses 
which originally led us to investigate analogous problems for the 
Dirac field, such as the quadrupole problem presented in this 
section and subsumed in (6.8) or (6.9), respectively. Accordingly, 
a comparison of Truedell's presentation of Signorini's theory with 
our equations (6.8) and (6.9) justifies thesis 3a.  

Thus the shear current ${\buildrel\frown\over\Upsilon\quer}{}^{ijk}$ 
of (4.3) can be considered as involved in exciting volume-preserving 
deformed-nuclei vibrational bands or hadronic Regge trajectories.
The space-integrals of its 9 components will constitute the 
algebraic generators of the coset space $SL(4,R)/SO(1,3)$ 
and can be derived from the invariance of the action under the linear 
group $SL(4,R)$, using the Noether theorem. Hence, together with 
the Lorentz generators and in accordance with thesis 3b, they generate 
the Lie algebra of the special linear group $SL(4,R)$. 

What is left then for study are the theses 4a and 4b. In other words, 
we have to turn our attention to the equilibrium conditions of contimuum 
mechanics (or rather statics) and to the conservation laws linked to 
spacetime symmetries in relativistic field theories, respectively. 
\bigskip\goodbreak

\sectio{\bf Conservation Laws for Momentum and Hypermomentum}
\medskip

We took the Dirac field as a concrete example for a field with an 
intrinsic spin current. If we develop the Lagrange-Noether formalism not 
in flat Minkowski spacetime, but in the Riemann-Cartan spacetime of the 
Poincar\'e gauge theory, then the laws (5.8) and (5.9) transmute into 
$$D\Sigma_{\alpha} =(e_{\alpha}\rfloor T^{\beta})\wedge\Sigma_{\beta}
+(e_{\alpha}\rfloor R^{\b\c})\wedge \tau_{\b\c}\eqa$$
and
$$D\tau_{\alpha\beta}+\vta_{[\alpha}\wedge\Sigma_{\beta]}=0\,,\eqa$$
respectively. Here $T^\a$ denotes the torsion (A.6) and $R_\a{}^\b$ 
the curvature (A.7) of spacetime. Clearly, the angular momentum law 
(7.2) remains the same, whereas the energy-momentum law (7.1) picks up 
Lorentz-type force densities on its right hand side. It is remarkable 
that the momentum law, if Minkowski spacetime is left, is supplemented 
by torsion and curvature dependent terms. However, if the spin 
vanishes, then (7.1) and (7.2) reduce to
$$\tau_{\a\b}=0\qquad\Longrightarrow\qquad D^{\{\}}\Sigma_\alpha =0 
\,, \qquad \vartheta_{[\a}\wedge\Sigma_{\b]}=0\,.\eqa$$
These equations describe the situation within the framework of GR and, 
if gravity is switched off, within SR. The description of ordinary 
macroscopic matter is achieved by a conserved and symmetric momentum 
current obeying (7.3). 

At a time when the deformation of strings, of membranes, and of 
higher extendons is still being thoroughly investigated for a possible 
application in elementary particle physics, it seems only natural 
to wonder whether the deformational properties and, linked 
therewith, the possibility for more complicated stress states of 
the spacetime continuum itself have been exhausted to a reasonable 
degree with (7.1) and (7.2). In fact, a comparison of (7.3) with the 
{\it special}-relativistic conservation laws of energy-momentum 
and angular momentum for a Dirac field, see (5.8) and (5.9), shows that 
not even the momentum and the spin distributions of a Dirac field 
can be accommodated in the classical picture of a continuum.
 
In Mindlin's micro-structured continuum, we have a {\it total} 
stress $\Sigma_a$ and a symmetric {\it Cauchy} stress $\sigma_
{ab}=\sigma_{ba}:=\vta_a\wedge\sigma_b$.
With the intrinsic hyperstress $\Delta^a{}_b$ of (3.10), 
Mindlin's equilibrium conditions in Cartesian coordinates read [27]:
$$D\Sigma_a=0\;,\qquad\qquad D\Bigl(\Delta^a{}_b+x^
a\wedge\Sigma_b\Bigr) =\sigma^a{}_b\,.\eqa$$
By differentiation and applying (7.4a), the last 
equation can be rewritten as 
$$D\Delta^a{}_b+\vta^a\wedge\Sigma_b=\sigma^a{}_b\,.\eqa$$
In the special case of {\it vanishing} hyperstress, we recover 
the classical equilibrium conditions:
$$\Delta^a{}_b=0\quad\Longrightarrow\quad D\Sigma_a=0\,,
\quad\vta_a\wedge\Sigma_b=\sigma_{ab}\quad\Longleftrightarrow\quad
\Sigma_a=\sigma_a\,,\quad\vta_{[a}\wedge\sigma_{b]}=0\,.\eqa$$

The structures which we find in evaluating Mindlin's theory of 
continua with microstructure, match exactly the concepts which arise 
if one supposes that spacetime obeys a metric-affine geometry with 
independend metric and connection. 

A metric-affine framework for a gravitational gauge theory has been set 
up in the literature, see [12, 16, 17, 18, 25] and, for the latest
developments, [35--38]. It is a generalization 
of GR in the same spirit, as Mindlin's theory of microstructured media 
is a generalization of classical elasticity. Consequently such 
theories are deeply rooted in concepts of differential geometry and 
continuum mechanics. In the 4 dimensions of spacetime, it is not too 
hard to find the 4-dimensional analogs of (7.4) and (7.5). Technically, 
one works out the gauge theory of the affine group 
$R^4\semidirect GL(4,R)$, in particular its Lagrange and Noether formalism. 
Then, we have to introduce the coframe $\vta^{\alpha}$, the 
connection $\Gamma_{\alpha}{}^{\beta}$ and the metric $g_{\alpha \beta}$
as gauge fields. The related field strengths are torsion $T^{\alpha}$,
curvature $R_{\alpha}{}^{\beta}$, and nonmetricity $Q_{\alpha \beta}$.
It turns out that the Noether identities are given by
$$D\Sigma_{\alpha} =(e_{\alpha}\rfloor T^{\beta})\wedge\Sigma_{\beta}
+(e_{\alpha}\rfloor R_{\beta}{}^{\gamma})\wedge \Delta^{\beta}{}_{\gamma}
-{1 \over {2}}(e_{\alpha}\rfloor Q^{\beta \gamma})\sigma_{\beta \gamma}
\eqa$$
and
$$D\Delta^{\alpha}{}_{\beta}+\vta^{\alpha}\wedge\Sigma_{\beta}=
g_{\beta \gamma}\,\sigma^{\alpha \gamma}\,,\eqa$$
respectively. They represent legitimate generalizations of the 
laws (7.1) and (7.2). They can be straightforwardly extended to n 
dimensions and justify the proposal of our thesis 4.
Obukhov and Tresguerres [35] developed the concepts of two different 
hyperfluids which both obey (7.7) and (7.8). Thus there exist suitable 
hydrodynamical representations of `hypermatter' which exactly fit the 
framework used in thesis 4. Therefore we can be sure that the hypermomentum 
concept is not empty and is non-trivial and represents a reasonable step in
the direction of a more refined description of matter.

\bigskip\goodbreak 
\centerline{\bf Acknowledgments}
\medskip
One of us is grateful to Professor Giorgio Ferrarese for the invitation to 
present an earlier version of this article as a lecture 
and for his hospitality on Elba. We thank 
Dermott McCrea${}^\dagger$ (Dublin) for help in improving the manuscript.
Some parts of the lecture are based on joint work with Dermott 
McCrea${}^\dagger$, Eckehard Mielke (Kiel), and Yuval Ne'eman (Tel Aviv).
\bigskip\goodbreak 
\centerline{\bf Appendix: On Exterior Calculus, Notations and Conventions}
\medskip
Let us start with a 4-dimensional differentiable 
manifold, an $X_4$ (compare Schou\-ten [31]). Its four 
coordinates we call $x^i$ (holonomic or coordinate indices 
$i,j,k... = 0,1,2,3$). At each point $P$ we have a tangent vector 
space $T_PX$. The tangent vectors $\partial_i$ along the 
coordinate lines form a vector basis. The corresponding basis of 
the dual vector space of 1-forms $T_P^*X$ will be denoted by $dx^j$, 
that is, $\partial_i\inner dx^j=\delta_i^j$, where we have 
introduced the interior product $\partial_i\inner dx^j \equiv dx^j
(\partial_i)$. 

We linearly transform the basis $\partial_i $. Thereby we introduce 
at each point $P$ an arbitrary vector basis (or frame) 
$ e_\alpha = e ^i {}_\alpha\,\partial_i $ (anholonomic or frame 
indices $\alpha,\beta,\gamma ... = 0,1,2,3$) and the dual 1-form basis 
(or coframe) $\vartheta ^\beta=e_j{}^\beta dx^j$, where $e_\alpha\inner 
\vartheta ^\beta = \delta _\alpha ^\beta $. The condition $e_\alpha\inner
\vartheta ^\beta=\delta_\alpha ^\beta$ determines the 1-forms 
$\vartheta ^\beta$ uniquely, since $e_j{}^\b$, as a basis, is 
non-degenerate. 
The coframe spans the algebra of tensor-valued exterior differential 
forms, see Choquet-Bruhat et al.[3] and Trautman [32]. 

For developing a physical theory, we 
need a tool for displacing a geometrical object from $P$ to a 
nearby point $P'$. Therefore we introduce a linear {\it 
connection}
$$\Gamma=\Gamma_\alpha{}^\beta\,L^\a{}_\b\,,\qquad\qquad\Gamma_
\alpha{}^\beta=\Gamma_{i\alpha}{}^\beta\,dx^i\;.\eqno(A.1)$$
The $L^\a{}_\b$ constitute a basis of the $gl(4,\Reel)$. Hence the 
connection $\Gamma$ is a 1-form  with values in the Lie algebra
$gl(4,\Reel)$. Then, for a tensor-valued p-form we can define the
$GL(4,\Reel)$-covariant exterior derivative $D$. It is given by 
$ D:= d +\Gamma _\alpha {}^\beta \,\rho(L^\alpha{}_\beta)\wedge$ , where 
$\rho$ denotes the representation type and the plus sign holds for an 
upper tensor index. On the coframe, for instance, $D$ acts according to   
$$ D \vartheta ^\alpha := d \vartheta ^\alpha + \Gamma _\beta {}^\alpha
   \wedge \vartheta ^\beta\;. \eqno(A.2) $$

The manifolds we are interested in carry, besides a connection, a 
Minkowskian {\it metric}. In an arbitrary coframe, the metric reads 
$g=g_{\alpha \beta}\,\vta^{\alpha}\otimes\vta^{\beta}$. It is always 
possible to choose an orthonormal basis wherein the metric reads 
$$ g=o_{\alpha \beta}{ \buildrel {orth.} \over {\vta^\alpha} }  \otimes
    { \buildrel {orth.} \over {\vta^{\beta} }} ,\qquad
   o_{\alpha \beta }:= diag(-1,1,1,1) \,.\eqno(A.3) $$
An $X_4$ with a linear connection and a metric we will call an $(L_4,g)$, 
a metric-affine space. 
                                       
Owing to the metric, we can define the invariant volume 4-form $ \eta $
by putting
$$ \eta  := { {1\over 4!}}  
   \eta_{\alpha \beta \gamma \delta } \, \vartheta ^\alpha 
   \wedge \vartheta ^\beta \wedge  \vartheta ^\gamma \wedge  \vartheta 
   ^\delta  =: \hodge{\bf 1}\,,\eqno(A.4)$$ 
with $ \eta _{\alpha \beta \gamma \delta } := 
\sqrt{|\hbox{det}(g_{\mu\nu })|} \, \epsilon_{\alpha \beta \gamma 
\delta }$, where $ \epsilon_{\alpha \beta \gamma \delta }$ is the totally 
antisymmetric Levi-Civita symbol, $\epsilon_{\hat 0 \hat 1 \hat 2 \hat 3} = 
+1$. Thereby we defined the action of the Hodge dual operator 
$\hodge$ on a 0-form. This operator is linear. Its action on the p-form 
basis $\vta^\a\wedge\vta^\b\wedge\cdots$ is defined below. It 
induces another representation of the bases which span the algebra 
of exterior forms on each $T^*_PX$:
\bea
\eta _\alpha  &:=&  {1\over 3!} \eta _{\alpha \mu \nu \rho}\, 
   \vartheta ^\mu \wedge \vartheta ^\nu \wedge \vartheta 
   ^\rho  = e_\alpha \inner \eta =: \hodge(\vartheta _\alpha ) \,,\nonumber \\
 \eta _{\alpha \beta } &:=&  
   {1\over 2!} \eta _{\alpha \beta \mu \nu}\,  
   \vartheta ^\mu \wedge  \vartheta ^\nu 
   = e_\beta \inner \eta _\alpha =:
   \hodge(\vartheta _\alpha \wedge \vartheta _\beta )  \,, \nonumber\\
 \eta _{\alpha \beta \gamma } &:=&  
   \eta _{\alpha \beta \gamma \mu }\,  
   \vartheta ^\mu = e _\gamma \inner \eta _{\alpha \beta } =:
   \hodge(\vartheta_\alpha\wedge\vartheta_\beta\wedge\vartheta_\gamma )
   \,, \nonumber \\
\eta _{\alpha \beta \gamma \delta } &=& e_\delta \inner 
   \eta_{\alpha \beta \gamma } =:
   \hodge(\vartheta_\alpha\wedge\vartheta_\beta\wedge\vartheta_\gamma 
   \wedge \vartheta _\delta )\,. \nonumber
\eea

\vspace*{-1.8cm}

$$
\eqno(A.5)
$$

In terms of the 1-forms $\vartheta^\alpha$, $\Gamma_\alpha{}^\beta$, 
and their exterior derivatives, we can define the {\it torsion} 2-form
$$T^\alpha:=D\vartheta^\alpha={1\over 2}\,T_{ij}{}^\alpha dx^i\wedge
   dx^j\eqno(A.6)$$
and the {\it curvature} 2-form
$$ R_\alpha {}^\beta := d \Gamma _\alpha {}^ \beta + \Gamma _\mu {}^\beta
   \wedge \Gamma _\alpha {}^\mu = {1\over 2}\, R_{ij\alpha }{}^\beta 
   dx^i \wedge dx^j\,. \eqno(A.7) $$
They  play the role of field strengths and the coframe and the 
connection that of potentials.

In general, the metric does not commute with the exterior 
covariant derivative. Hence by differentiation, as in the case of 
coframe and connection, we can define the field strength
attached to the metric. We call it the {\it nonmetricity} 1-form:
$$ Q_{\alpha \beta } := -D g_{\alpha \beta } = Q_{i\alpha \beta }
   \; dx^i \,. \eqno(A.8) $$
The Weyl 1-form (or Weyl covector) is given by the trace
$$ Q:= {1\over 4} g^{\alpha \beta } Q_{\alpha \beta }  \,. \eqno(A.9) $$

We collect in a table the potentials and the field strengths as well 
as the Bianchi identities of a metric-affine space $(L_4,g)$, see [17]:
$$\vbox{\offinterlineskip
\hrule
\halign{&\vrule#&\strut\quad\hfil#\quad\hfil\cr
height2pt&\omit&&\omit&&\omit&\cr
& {\bf potential } && {\bf field strength } && {\bf Bianchi identity } &\cr
height2pt&\omit&&\omit&&\omit&\cr
\noalign{\hrule}
height2pt&\omit&&\omit&&\omit&\cr
& metric $g_{\alpha \beta }$ && $Q_{\alpha \beta }= -Dg_{\alpha \beta }$ && 
$DQ_{\alpha \beta }= 2 R_{(\alpha \beta )}$ &\cr 
& coframe  $\vartheta ^\alpha $  && 
$T^\alpha = D\vartheta ^\alpha $ && $DT^\alpha = R_\mu {}^\alpha \wedge 
\vartheta ^\mu $ &\cr
& connection $ \Gamma _\alpha {}^\beta $ && $ R_\alpha {}^\beta 
= d \Gamma _\alpha {}^\beta + \Gamma _\mu {}^\beta \wedge 
\Gamma _\alpha {}^\mu $ && $D  R_\alpha {}^\beta =0 $ &\cr 
height2pt&\omit&&\omit&&\omit&\cr}
\hrule}$$
\medskip
For $Q_{\a\b}=0$, we find the Riemann-Cartan space $U_4$ and, if also 
$T^\a=0$, we recover the Riemannian space $V_4$ of general relativity. 
Flat Minkowski space $M_4$ is obtained if, additionally, 
$R_{\alpha}{}^{\beta} =0$.

Up to here, we presented the 4-dimensional formalism of general-relatistic 
field theory. Of course, for {\it three dimensions}, we have the analogous 
structures and quantities. The local metric will be $o_{\rho\sigma}
={\rm diag}(1,1,1)$ and the $\eta$-basis will be restricted to 
$(\eta,\,\eta_\rho,\,\eta_{\rho\sigma},\,\eta_{\rho\sigma\tau})$. We 
denote the (holonomic) coordinate indices in 3 dimensions by 
$a,b,c ...=1,2,3$, the (anholonomic) frame indices by $\rho,\sigma,
\tau ...=1,2,3$. 

In ordinary continuum mechanics, see Truesdell [34], the underlying manifold 
describing the continuum to be deformed is Euclidean. However, if we 
admit the distribution of dislocations, for example, then the underlying 
manifold is a flat $U_3$, see Kr\"oner [21]. Therefore also in the 
3-dimensional case the field strengths listed in our table can be 
non-vanishing and may play an essential role in characterizing a 
continuum. 
\pagebreak

\centerline{\bf References}
\newref [1]  W.L.~Burke: {\it Applied differential geometry} 
          (Cambridge University Press, Cambridge 1985).
\newref [2] G. Capriz, {\it Continua with Microstructure} (Springer,
Berlin 1989). 
\newref [3]  Y.~Choquet-Bruhat, C.~DeWitt-Morette, and M.~Dillard-%
Bleick: {\it Analysis, Manifolds and Physics}, rev.\ ed. 
(North Holland, Amsterdam 1982).
\newref [4] E. et F. Cosserat, {\it Th\'eorie des corps d\'eformables}
(Hermann et Fils, Paris 1909).
\newref [5] C.T.J.~Dodson and T.~Poston, {\it Tensor Geometry}, 2nd ed.
(Springer, Berlin 1991). 
\newref [6] P.~Dombrowski (Cologne), private communication (1992).
\newref [7] Y. Dothan, M. Gell-Mann, and  Y. Ne'eman, Phys. Lett. {\bf 17},
148 (1965).
\newref [8] J.L.~Ericksen, Arch. Rat. Mech. Anal. {\bf 113}, 97 (1991).
\newref [9] A.C.~Eringen and C.B.~Kafadar, in {\it Continuum Physics}, 
A.C.~Eringen (ed.), Vol.IV  (Academic Press, New York 1976) p.1.  
\newref [10] G.~Ferrarese, Ann. Mat. Pura Appl. {\bf 108}, 109 (1976).
\newref [11] G.Grioli, {\it Mathematical theory of elastic equilibrium} 
 (Springer, Berlin 1962).  
\newref [12] R. Hecht and F.W.~Hehl, in {\it General Relativity and 
Gravitational Physics}, Capri 1990, R.Cianci et al.\ (eds.) 
(World Scientific, Singapore 1991) p.246.  
\newref [13]  F.W.~Hehl, Reports on Math.~Phys. (Torun) {\bf 9}, 55 (1976).
\newref [14]  F.W.~Hehl, J. Lemke, and E. W. Mielke, in {\it Geometry and 
Theoretical Physics}, Bad Honnef 1991, J.~Debrus and A.C.~Hirshfeld 
(eds.) (Springer, Berlin 1991) p.56. 
\newref [15] F.W.~Hehl and J.D.~McCrea, Found. Phys. {\bf 16}, 267 (1986).
\newref [16] F.W.~Hehl, E.A.~Lord, and Y. Ne'eman, Phys. Rev. {\bf D17},
428 (1978).
\newref [17] F.W.~Hehl, E.W.~Mielke, J.D.~McCrea, and Y.~Ne'eman,
Found.~Phys. {\bf 19}, 1075 (1989); Physics Reports {\bf 258} 1 (1995). 
\pagebreak
\newref [18] F.W.~Hehl and Y. Ne'eman, in {\it Modern Problems of 
Theoretical Physics (Fest\-schrift for D. Ivanenko)}, P.I. Pronin and Yu.N. 
Obukhov (eds.) (World Scientific, Singapore 1991) p.31 
\newref [19] W.~Jaunzemis, {\it Continuum Mechanics} (Macmillan, New York 
1967).
\newref [20] W.~Kopczy\'nski and A.~Trautman, {\it Spacetime and Gravitation}
(PWN, Warszawa, and Wiley, Chichester 1992).  
\newref [21] E.~Kr\"oner, in {\it Physics of Defects},
          Les Houches, Session XXXV, 1980. R. Balian et al. (eds.)
          (North-Holland, Amsterdam 1981) p.215.
\newref [22]   E.A. Lord and P. Goswami, J. Math. Phys.  
{\bf 27}, 2415 (1986). 
\newref [23] G.A.~Maugin and A.C.~Eringen. J. Math. Phys. {\bf 13}, 1788
(1972).
\newref [24] J.D.~McCrea, in: {\it Proceedings of the 14th Int.~Conference on
          Differential Geometric Methods in Mathematical Physics}, Salamanca
          1985, P.L.~Garc\'ia and A.~P\'erez-Rend\'on (eds.), Lecture Notes
          in Mathematics (Springer), Vol. {\bf 1251} (1987) p.222.
\newref [25] J.D.~McCrea, F.W.~Hehl, and E.W.~Mielke,
          Int.~J.~Theor.~Phys. {\bf 29}, 1185 (1990).
\newref [26] E.W.~Mielke, {\it Geometrodynamics of Gauge Fields} 
(Akademie-Verlag, Berlin 1987).
\newref [27]  R.D.~Mindlin, Arch.~Rat.~Mech.~Anal. {\bf 16}, 51 (1964).
\newref [28] Y.~Ne'eman and Dj.~\v Sija\v cki, Phys. Lett. {\bf B157},
275 (1985); (E) {\bf B160}, 431 (1985).
\newref [29] Y.~Ne'eman and Dj.~\v Sija\v cki, Phys. Rev. {\bf D37}, 
3267 (1988).
\newref [30] Y.~Ne'eman and Dj.~\v Sija\v cki, Phys. Lett. {\bf B200},
489 (1988).
\newref [31]  J.A.~Schouten, {\it Ricci-Calculus}, 2nd   
ed. (Springer, Berlin 1954).
\newref [32] A.~Trautman, in: {\it Differential Geometry, Symposia Matematica}
          {\bf 12} (Academic Press, London 1973) p.139.
\newref [33] C.~Truesdell and R.A.~Toupin: {\it The Classical Field Theories},
          in: {\it Handbuch der Physik} (S.~Fl\"ugge ed.), Vol. III/1,
           (Springer, Berlin 1960) p.226.
\newref [34] C.~Truesdell, {\it A First Course in Rational Continuum 
Mechanics I} (Academic Press, New York 1977).
\pagebreak

\newref [35] Yu.N. Obukhov and R. Tresguerres, Phys. Lett. {\bf A184}, 17
     (1993); Yu.N. Obukhov, Phys. Lett. {\bf A210}, 163 (1996).

\newref [36] F. Gronwald and F.W. Hehl, in: {\it Proc.\ of the 14th Course of
     the School of Cosmology and Gravitation on `Quantum Gravity',}
     held at Erice, Italy, May 1995, P.G. Bergmann, V. de Sabbata, and
     H.-J. Treder, eds. (World Scientific, Singapore 1996) p.\ 148.

\newref [37] E.J. Vlachynsky, R. Tresguerres, Yu.N. Obukhov, and F.W. Hehl,
     Class. Quantum Grav. {\bf 13}, 3253 (1996).

\newref [38] T. Dereli, M. \"Onder, J. Schray, R.W. Tucker, and C. Wang,
     Class.  Quantum Grav. {\bf 13}, L103 (1996).

\bigskip
\centerline{==================}

\end{document}